\documentclass[aps,pra,epsfigure,twocolumn,showpacs]{revtex4}
\usepackage{dcolumn}
\usepackage{bm,enumerate}
\usepackage{graphicx}
\usepackage{amsmath}
\usepackage{latexsym}
\usepackage{amsfonts}
\usepackage{amssymb}
\usepackage{array}
\usepackage{epstopdf}
\usepackage{txfonts}

\newcommand{\ket}[1]{\left\vert#1\right\rangle}

\newcommand{\bra}[1]{\left\langle#1\right\vert}

\newcommand{\up}{\uparrow}
\newcommand{\down}{\downarrow}

\begin{document}
\title{Propagation of non-classical correlations across a quantum spin chain}
\author{S. Campbell$^{1,2,3}$, T. J. G. Apollaro$^4$, C. Di Franco$^2$, L. Banchi$^{4,5}$, A. Cuccoli$^{4,5}$, R. Vaia$^6$, F. Plastina$^{7,8}$, and
 M. Paternostro$^1$}
\affiliation{$^1$Centre for Theoretical Atomic, Molecular and Optical Physics, School of Mathematics and Physics, Queen's University, Belfast BT7 1NN, United Kingdom\\
$^2$Department of Physics, University College Cork, Republic of Ireland\\
$^3$Quantum Systems Unit, Okinawa Institute of Science and Technology, Okinawa, Japan\\
$^4$Dipartimento di Fisica e Astronomia, Universit\`a di Firenze,~Via G. Sansone 1, I-50019 Sesto Fiorentino (FI), Italy\\
$^5$INFN Sezione di Firenze, via G.Sansone 1, I-50019 Sesto Fiorentino (FI), Italy\\
$^6$Istituto dei Sistemi Complessi, Consiglio Nazionale delle Ricerche, via Madonna del Piano 10, I-50019 Sesto Fiorentino (FI), Italy\\
$^7$Dipartimento di  Fisica, Universit\`a della Calabria, 87036 Arcavacata di Rende (CS), Italy\\
$^8$ INFN - Gruppo collegato di Cosenza, Universit\`a della Calabria, 87036 Arcavacata di Rende (CS), Italy}

\date{\today}

%\date{\today}
%ABSTRACT
\begin{abstract}
We study the transport of quantum correlations across a chain of
interacting spin-$1/2$ particles. As a quantitative figure of
merit, we choose a symmetric version of quantum discord and
compare it with the transported entanglement, addressing various
operating regimes of the spin medium. Discord turns out to be
better transported for a wide range of working points and initial
conditions of the system. We relate this  behavior to the
efficiency of propagation of a single excitation across the spin
chain. Moreover, we point out the role played by a magnetic field in 
the dynamics of discord in the effective 
channel embodied by the chain. Our analysis can be interestingly extended to transport
processes in more complex networks and the study of non-classical correlations under 
general quantum channels.
\end{abstract}
\maketitle

The behavior of features such as quantum coherence and entanglement in a composite quantum system 
whose state is exposed to the effects of environmental actions has been the focus of an extensive research activity. 
Recently, much attention has been paid to the case of environments embodied by systems of interacting quantum
particles~\cite{amico}. Such dynamical environments can induce interesting back actions on the evolution of a system,
thus significantly affecting its properties. From the point of view of coherent information processing, on the other hand, the 
non-trivial dispersion properties of networks of such interacting particles represent an interesting opportunity for their use as 
short-haul communication channels for the inter-connections among on-chip nodes in the next generation of information processing devices~\cite{bosereview}. 

While most of the work in these contexts has focused on the study of the properties of entanglement upon propagation in such media, it is now widely accepted that the space of non-classical correlations accommodates more than {\it just} quantum entanglement. 
Figures of merit such as quantum discord~\cite{discord1,discord2} and measurement-induced disturbance~\cite{Luo2008}, to cite only two of the most popular ones, are able to capture the content of non-classical correlations of a state well beyond entanglement. Although the role played by such broader forms of non-classical correlations in the quantum mechanical manipulation of information has yet to be fully understood, enormous is the interest they bring about as the manifestation of the various facets of quantumness in a system. It is thus very important to work on the exploration of the behavior of such quantities upon exposure to dynamical and finite environments of the sort addressed above, so as to build a useful parallel with the much more extensively investigated case of entanglement.

In this paper we study the propagation of quantum correlations
across a system of interacting spin-$1/2$ particles. Our main goal
is to compare the way important indicators of non-classicality,
such as quantum discord (QD)~\cite{discord1,discord2} and
entanglement of formation (EoF)~\cite{wootters}, are transferred
through a medium offering non-trivial dispersion properties. In
doing this, we aim at understanding whether or not the
fundamentally conceptual difference between entanglement and
discord leaves signatures in the way such non-classical quantities
are transferred. We show that this is indeed the case by preparing
a non-separable (in general mixed) state of an isolated spin and
the one occupying the first site of a linear spin-chain. We then
compare the quantum-correlation properties of such an initial state
with those of the state achieved, at a given instant of time of
the evolution, between the isolated spin and the one occupying the last site of the
chain itself. QD appears to be better transmitted than
entanglement (as quantified by EoF) in a wide range of working
conditions and regardless of the details of the initial state
being considered. It is more robust to the dispersion inherent in
the effective spin-medium across which it propagates, being
non-zero in situations where the EoF is, for all practical
purposes, null. By relating the entanglement to the
single-excitation transition amplitude of the system, we identify
the working point at which a cross-over occurs between the quality of transport of QD and EoF,
making the transport of entanglement more efficient. Moreover,
interesting effects of entanglement-forerunning~\cite{davidovich},
where QD {\it precedes} the establishment of EoF, are found in the
way quantum correlations build up between the isolated spin and the last one
in a given chain. Our analysis considers a large number of state
families in such a transport problem, addressing explicitly those
that maximize the degree of discord at given global
mixedness~\cite{alqasimi, MNCMS,zambrini}.

Our study provides exact quantitative answers to a problem that
has been so far largely overlooked, although being relevant for a
wide range of physical situations. For instance, it is sufficient
to think about recent studies of the propagation of information in
biological systems operating on the verge of quantumness, which
appear to benefit from the inclusion of a mild degree of
noise~\cite{bio}. In such conditions, one might wonder whether
other forms of quantum correlations are favored, given that the
perfect transport of entanglement would be prevented.

The remainder of this paper is organized as follows. In
Sec.~\ref{model} we describe the physical situation at hand and
provide the general analytical form of the time-dependent 
density matrix describing the state of the isolated spin and the last one in a chain. 
Sec.~\ref{tools} is devoted to a brief
introduction to the quantitative indicators of quantum
correlations adopted in this work. In Sec.~\ref{Ss.case} we face
the propagation of EoF and QD across the spin chain and address
the relation between such non-classicality indicators and the
single-excitation transition amplitude. While the transport of
discord appears to be favored, we point out the existence of a
cross-over point in the parameter space at which the performance
of EoF becomes superior. Our study includes both pure and mixed
input states, among them the case embodied by states
that maximize QD at fixed global entropy.
%, which are thus expected
%to optimize the resilience of discord under the transport process.
Sec.~\ref{Ss.magnetic} addresses the effect of a uniform magnetic
field on the dynamics of the discord finding that it has
relevance only for states lacking of rotational symmetry in the
$XY$-spin plane and, in Sec.~\ref{Ss.environment}, we describe
briefly the evolution of the non-classicality indicators under the
influence of an environment modelled by a spin chain. Finally, in
Sec.~\ref{conc} our conclusions are drawn and some open questions
arising from the present work are put forward.

\section{The Model}
\label{model}

We consider the configuration shown in Fig.~\ref{sketch}, {\textit{i.e.},
a quantum channel consisting of $N$ interacting spin-${1}/{2}$ particles
in a linear configuration with open boundary conditions. The Hamiltonian model describing the
system is taken to be (we take units such that $\hbar\,{=}\,1$ throughout the manuscript)
\begin{equation}
\label{e.channelham}
\hat{\cal H}=
 -2J \sum_{i=1}^{N-1}(\hat S^i_x \hat S^{i+1}_x
 +\hat S^i_y \hat S^{i+1}_y)
 {-}2 h\sum_{i=1}^{N} \hat S^i_z
\end{equation}
with $\hat S_k^i$ the $k\,{=}\,x,y,z$ spin-component operator of particle
$i\,{=}\,1,...,N$, $J$ the inter-spin coupling strength and $h$ a
uniform magnetic field. In what follows, the spins occupying sites $j\,{=}\,2,...,N$ will
be assumed to be all prepared in down state $\ket{\down}$, with
$\{\ket\down,\ket\up\}$ denoting the eigen-states of $\hat S_z$.
On the other hand, the first spin forms a (generally mixed)
bipartite quantum correlated state with a further particle,
labelled $0$, which is physically detached from the chain.

\begin{figure}[t]\label{F.schemeMauro}
\includegraphics[%trim= 0mm 55mm 0mm 50mm, clip, 
width=0.9\linewidth]{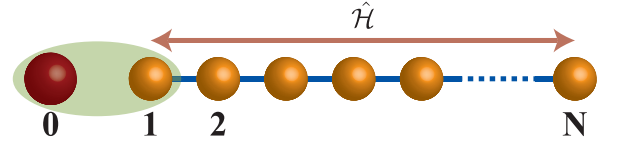}
\caption{(Color online) We consider a chain of $N$ interacting
spin-$1/2$ particles coupled through the Hamiltonian model
$\hat{\cal H}$ [cf. Eq.~(\ref{e.channelham})]. A further
spin-$1/2$ particle, labeled $0$ and completely isolated from the
chain, is prepared in a joint mixed state $\rho^{(1,0)}(0)$ with
particle $1$. We study how the general quantum correlations of
such a state propagate across the chain.} \label{sketch}
\end{figure}
The symmetry properties of the Hamiltonian model in Eq.~\ref{e.channelham} restrict 
the dynamics to those states within the zero- and
the single-excitation sectors of the total Hilbert space of the spins~\cite{son}, where the
chain behaves as an amplitude damping channel~\cite{bose} fully
characterized by the transition amplitude of the spin excitation
from site $1$ to site $r\,{=}\,1,...,N$. The latter is conveniently
expressed as $f_{r}(t)\,{=}\,\bra{r} e^{-i\hat{\cal H} t}\ket{1}$,
where in the states $\ket{n}$ ($n\,{=}\,1,...,N$) all of the spins
are in $\ket{\downarrow}$ except the one at position $n$, which is
in $\ket{\uparrow}$.

In this basis, $\hat{\cal H}$ is represented by an $N\times N$
tridiagonal matrix, which can be analytically diagonalized for any
length. The reduced density matrix $\rho^{(r)}(t)$ describing the
state of particle $r$ at time $t$ can thus be expressed as a
function of the state of particle $1$ at the reference time
$t\,{=}\,0$. That is
\begin{equation}
 \rho^{(r)}(t)=\left[\begin{matrix}
 \rho^{(1)}_{\down\down}(0)+\rho^{(1)}_{\up\up}(0)(1-|f_{r}(t)|^2) &\rho^{(1)}_{\down\up}(0) f_{r}(t) \\
  \rho^{(1)}_{\up\down}(0) f_{r}^*(t)& \rho^{(1)}_{\up\up}(0)|f_{r}(t)|^2
\end{matrix}\right],
\end{equation}
where
$\rho_{\alpha\beta}=\bra{\alpha}\rho\ket{\beta}~(\alpha,\beta\,{=}\,\down,\up)$.

In the remainder of this work we will consider the chain as prepared in the fully factorized 
state $\otimes^N_{j=2}\ket{\down}_j$, while the initial state of spins $0$ and $1$ is a quantum 
correlated state. Clearly, while spin $0$ undergoes only a free evolution, the chain's elements 
evolve according to the intra-chain coupling model in Eq.~(\ref{e.channelham}). This implies that the 
overall time evolution operator $\hat{\cal U}(t)$ that propagates the state of the $N+1$ spins factorizes as
$\hat{\cal U}(t)=\hat{\openone}\otimes{e}^{-i\hat{\cal H}t}$. Together with the uncorrelated initial state of the 
remaining part of the chain (which thus shares no correlation with spins $0$ and $1$), this legitimately allows us to 
make use of the formalism developed in Ref.~\cite{Bellomo2007} to get the 
joint state $\rho^{(r,0)}(t)$ of spins $r$ and $0$ from the knowledge of $\rho^{(r)}(t)$. 
Notice also that a very similar approach, which has also been verified by an exact numerical study,
 has been previously used in similar contexts~\cite{bose0}.
Although such an approach is broadly valid and can indeed be used for any initial state of spins $0$ and $1$,
here we restrict our analysis to $X$-type input states of the general form
\begin{equation}
\label{XXX}
\rho^{(1,0)}(0)=\left[\begin{matrix}
 \rho_{11} & 0 & 0 & \rho_{14} \\
0 & \rho_{22} & \rho_{23} & 0 \\
0 & \rho^*_{23} & \rho_{33} & 0 \\
 \rho^*_{14} & 0 & 0 & \rho_{44}
\end{matrix}\right]~~~\text{with}~\sum^4_{j=1}\rho_{jj}{=}1.
\end{equation}
Here we have introduced the compact notation $\rho_{ij}{=}\langle{i}|\rho^{(1,0)}(0)\ket{j}$ with $\ket{1} =\ket{ \downarrow \downarrow} ,\ket 2 =
\ket{\downarrow \uparrow} , \ket{3} = \ket{\uparrow \downarrow }, \ket 4 = \ket{\uparrow
\uparrow}$. In fact, as mentioned above, the excitation-preserving nature of the Hamiltonian
studied here (which commutes with the total number of excitations
in the system) ensures that the $X$-type character of any input state of the pair $(1,0)$  is preserved upon evolution.
%one of the two-spin computational states $\ket{\alpha,\beta}$,
%the evolved state will be of the form in Eq.~(\ref{XXX}) and the decoherence function will be the same
More explicitly, the only non-zero elements of the evolved density matrix will be
%\begin{widetext}
\begin{equation}
\begin{aligned}
\rho^{(r,0)}_{11}(t)&\,{=}\,\rho_{11}\,{+}\,(1\,{-}\,|f_{r}(t)|^2)\rho_{33}\,\,\,,\,\,\,\rho^{(r,0)}_{33}(t)\,{=}\,|f_{r}(t)|^2\rho_{33},\\
%&+(1-|f_{r}_{r's'}(t)|^2)(1-|f_{r}_{rs}(t)|^2)\rho_{44},\\
\rho^{(r,0)}_{22}(t)&\,{=}\,\rho_{22}\,{+}\,(1\,{-}\,|f_{r}(t)|^2)\rho_{44}\,\,\,,\,\,\,\rho^{(r,0)}_{44}(t)\,{=}\,|f_{r}(t)|^2\rho_{44},\\
%\rho^{(r0)}_{44}(t)=|f_{r}(t)|^2\rho_{44},\\
\rho^{(r,0)}_{14}(t)&\,{=}\,f_{r}(t) \rho_{14}\,\,\,,\,\,\,
\rho^{(r,0)}_{23}(t)\,{=}\,f_{r}(t) \rho_{23}.
\end{aligned}\label{e.rhooutrr'}
\end{equation}
The initial conditions being specified by the input state, the
above equations describe both the propagation of quantum
correlations from site 1 to site $r$, when $r\,{>}\,1$, and the
decohering influence of an environment (embodied by the spin
chain) on the spin occupying site 1, when $r\,{=}\,1$. In the
former case, the single-excitation transition amplitudes
$f_{r}(t)$ contain information on the working conditions of the
channel. The limiting case  $f_{r}(t)\,{=}\,1$ gives
$\rho^{(r,0)}(t)\,{=}\,\rho^{(1,0)}(0)$, {\it i.e.}, the perfect
{\it transfer} of the input state from pair $(1,0)$ to $(r,0)$. On
the other hand, $f_{1}(t)$ defines the probability amplitude of
finding the excitation on the first spin. Note that X-type density
matrices are such that no single-spin coherence will develop in
time, {\it i.e.}, both the reduced single-spin density matrices
remain diagonal.

\section{Figures of merit for quantum correlations}
\label{tools}

After having introduced the dynamical model that will be addressed
in our study, we turn our attention to the figures of merit that
will be used in order to perform our quantitative analysis. As
already anticipated, we take QD~\cite{discord1,discord2} as a
measure for general quantum correlations between any two spins under study. As originally proposed by Ollivier and Zurek, QD can be
associated with the difference between two classically equivalent
versions of mutual information, which measures the total
correlations within a quantum state. For a two-spin state
$\rho^{(r,r')}$ extracted from our system, the mutual information
is defined as ${\cal I}(\rho^{(r,r')})\,{=}\,{\cal
S}(\rho^{(r)}){+}{\cal S}(\rho^{(r')}){-}{\cal S}(\rho^{(r,r')})$.
Here, ${\cal S}(\rho)\,{=}\,{-}\text{Tr}[\rho\log_2\rho]$ is the
von Neumann entropy of a generic state $\rho$. Alternatively, one
can consider the one-way classical correlation ${\cal
J}^\leftarrow(\rho^{(r,r')})\,{=}\,{\cal S}(\rho^{(r)}){-}{\cal
H}_{\{\hat\Pi_i\}}(r|r')$~\cite{discord2}, where we have
introduced ${\cal H}_{\{\hat\Pi_i\}}(r|r'){\equiv}\sum_{i}p_i{\cal
S}(\rho^i_{r|r'})$ as the quantum conditional entropy associated
with the the post-measurement density matrix
$\rho^i_{r|r'}\,{=}\,\text{Tr}_{r'}[\hat\Pi_i\rho^{(r,r')}]/p_i$
obtained upon performing the complete projective measurement
$\{\Pi_i\}$ on spin $r'$
($p_i\,{=}\,\text{Tr}[\hat\Pi_i\rho^{(r,r')}]$). QD is thus
defined as
\begin{equation}
{\cal D}^\leftarrow\,{=}\,\inf_{\{\Pi_i\}}[{\cal I}(\rho^{(r,r')}){-}{\cal J}^\leftarrow(\varrho^{(r,r')})]
\end{equation}
with the infimum calculated over the set of projectors
$\{\hat\Pi_i\}$ \cite{discord1}. Analogously, one can define
${\cal D}^\rightarrow$, which is obtained upon swapping the roles
of $r$ and $r'$. The inherently asymmetric definition of QD makes,
quite naturally, ${\cal D}^\rightarrow\,{\neq}\,{\cal
D}^\leftarrow$. This might be the cause of misinterpretations: a
{\it quantum-classical} state for which ${\cal
D}^\leftarrow\,{\neq}\,0$ but ${\cal D}^\rightarrow\,{=}\,0$ (or
viceversa)~\cite{piani} might be interpreted as strictly classical
if only ${\cal D}^\rightarrow$ (${\cal D}^\leftarrow$) is probed.
Here, we are interested in the transport of quantum correlations,
regardless of the way they are encoded in the two-spin state.
Rather refined solutions to this issue passing through the
generalization of the definition of QD or the introduction of
strictly faithful entropic measures that are null only for
classical-classical states ({\it i.e.}, states such that ${\cal
D}^{\leftarrow,\rightarrow}\,{=}\,0$) have been proposed
~\cite{MNCMS}. However, they typically require a double
optimization to be performed over a bilateral set of projective
measurements. In order to bypass the numerical burden that this
would imply, we consider the two-way QD,
\begin{equation}\label{e.Discord}
\mathcal{D}\,{=}\,\text{max}[\mathcal{D}^\leftarrow, \mathcal{D}^\rightarrow],
\end{equation}
which is strictly null only on states endowed with no quantum
correlations, and faithfully signals classical-classical
states~\cite{piani}.

On the other hand, our chosen entanglement measure is
EoF~\cite{wootters}, which quantifies the minimum number of Bell
pairs needed in order to prepare a copy of the state
$\rho^{(r,r')}$ we are studying. The relationship between EoF and
QD has been recently examined to study the distribution of quantum
correlated states in the entropic space~\cite{alqasimi}. Moreover,
it is possible to establish a { triangular} relation connecting
QD,  EoF and conditional entropy in multi-spin quantum
states~\cite{ChinaBrasil}, so that such two figures of merit
appear to be natural choices for a quantitative comparison. For
arbitrary two-spin states, EoF is calculated as
\begin{equation}
\mathcal{E}\,{=}\,h\left(\frac{1}{2}\left[1+\sqrt{1-C^2}\right]\right)
\end{equation}
where $h(x)\,{=}\,-x\text{log}_2 x - (1-x)\text{log}_2 (1-x)$ is
the binary entropy function and $C$ is the concurrence of the
state~\cite{wootters}. The latter, an equally valid entanglement
measure, is found in terms of the eigenvalues $\lambda_1\,{\geq}\,
\lambda_{2,3,4}$ of the matrix
$\rho^{(r,r')}(\hat\sigma_y\otimes\hat\sigma_y)\rho^{(r,r')*}
(\hat\sigma_y\otimes\hat\sigma_y)$ as
\begin{equation}
C\,{=}\,\text{max}\left[0,\sqrt{\lambda_1}-\sum_{i=2}^{4}\sqrt{\lambda_i}\right],
\end{equation}
where $\hat\sigma_y$ is the $y$-Pauli operator. For an $X$-type state of pair $(r,r')$,
the concurrence is straightforwardly shown to be
\begin{equation}\label{e.conc2}
C_{(r,r')}=2 \max\left[0,\left|\rho_{14}\right|-\sqrt{\rho_{22} \rho_{33}},\left|\rho_{23}\right|-\sqrt{\rho_{11} \rho_{44}}\right].
\end{equation}
Differently from EoF, QD does not have a closed analytical
expression for any two-spin state, although some steps toward this goal have been performed~\cite{Ali2010,Girolami2011}.
Nevertheless, for the special case of the class of states presented
in Eq.~(\ref{XXX}), one can obtain analytic formulas for ${\cal
D}$ as a function of the dynamical parameter $f(t)$, as well as of
the input state. However, as their expressions are lengthly and not very informative, we
do not report them explicitly here.

\section{Propagation Of Quantum Correlations}
\label{single}

\subsection{Case Study of Pure States and MDMS}
\label{Ss.case} We perform our analysis by addressing the
propagation of quantum correlations across the spin chain when the
pair of spins $(1,0)$ is initialized in a given non-separable
state, while the remaining spins are in
$\otimes^N_{j=2}\ket{\down}_j$. In order to address the temporal
behavior of our figures of merit, we need the explicit form taken
by the single-excitation transition amplitude $f_{r}(t)$. Although
the present formalism is valid without any major difficulties for
general $r$ belonging to the chain, for the sake of clarity we set
hereafter $r\,{=}\,N$ and omit the subscript in $f(t)$. For a
uniform chain ruled by Eq.~(\ref{e.channelham}) such quantity reads
% \begin{equation}
% \label{E.amplitude}
% f(t)=\frac{2}{N+1}\sum^N_{k=1}\sin\left(\frac{k\pi}{N+1}\right)
% \sin\left(\frac{k\pi{N}}{N+1}\right)\text{e}^{-2it\left(h+\cos\frac{k\pi}{N+1}\right)}.
% \end{equation}
\begin{equation}
\label{E.amplitude}
f(t)=\frac{2}{N+1}\sum^N_{k=1}\sin{\textstyle\frac{k\pi}{N{+}1}}
\sin{\textstyle\frac{k\pi N}{N{+}1}}\text{e}^{-2it\left(h+\cos\frac{k\pi}{N+1}\right)}.
\end{equation}
We will use this explicit result to build up the state of spins $0$ and $N$ and thus calculate quantum correlations. The single-excitation transition amplitude 
is a real (purely imaginary) quantity for $N$ odd (even), due to the symmetry properties of the spectrum of the system 
(a more detailed discussion on the properties of function $f(t)$ is given in Ref.~\cite{Banchi2010}).
%where also non-homogeneous end-point couplings in Eq.~\ref{e.channelham} have been taken into account in order to maximize $f(t)$.

As a first significant instance, we consider the case in which the
joint state of spins $0$ and $1$ is pure. In fact, for pure states
${\cal E}={\cal D}$ and it is interesting to study whether or not,
in this case, entanglement is lost in favor of discord as
information propagates across our dispersive medium. As the EoF is
based on concurrence, we consider the
class of pure entangled input states parameterized as
\begin{equation}
\label{purestate}
\ket{\psi(C_{(1,0)})}_{(1,0)}=\sin\gamma\ket{\down\down}+\cos\gamma\ket{\up\up},
\end{equation}
where $\gamma{=}({1}/{2})\arcsin C_{(1,0)}$.
The evolved state is then achieved using the approach outlined in Sec.~\ref{model}.

Perfect transfer of entanglement across an interacting spin chain
is known to occur under proper conditions~\cite{kay}. In
particular, it was proved that perfect entanglement transfer is
achieved  when perfect end-to-end state transfer is made possible.
For a system of three spins, the Hamiltonian model in Eq.~\ref{e.channelham}
allows for perfect state transfer~\cite{cambridge}, thus implying
that this is also the case for entanglement and QD, since for pure
states ${\cal E}\,{=}\,{\cal D}$. Nonetheless, the examination of
Fig.~\ref{figure1} shows some interesting features. Clearly, the
peaks shown in all panels of Fig.~\ref{figure1} are achieved at
the instants of time at which the input state \eqref{purestate} is
perfectly transmitted, where the pair $(3,0)$ is pure and ${\cal
E}\,{=}\,{\cal D}\,{=}\,{\cal E}_{(1,0)}$. However, between two
consecutive peaks, the state of spins $0$ and $N$ is mixed and the
two figures of merit can be quantitatively different. Evidently,
within these time windows the transport of QD is favored with
respect to EoF, being not only quantitatively larger than ${\cal
E}$ but also non-null at times such that the EoF is, for all
practical purposes, zero. Interestingly, as we increase the
initial degree of entanglement of $\ket{\psi(C_{(1,0)})}_{(1,0)}$,
this effect becomes less important until, at $C_{(1,0)}\simeq1$,
there are narrow regions close to the peaks where the transported
entanglement overcomes QD. We will see later on in this
paper how this effect depends on the
single-excitation transition amplitude .

\begin{figure}[t]
{\bf (a)}\hskip3cm{\bf (b)}
\includegraphics[width=0.5\linewidth]{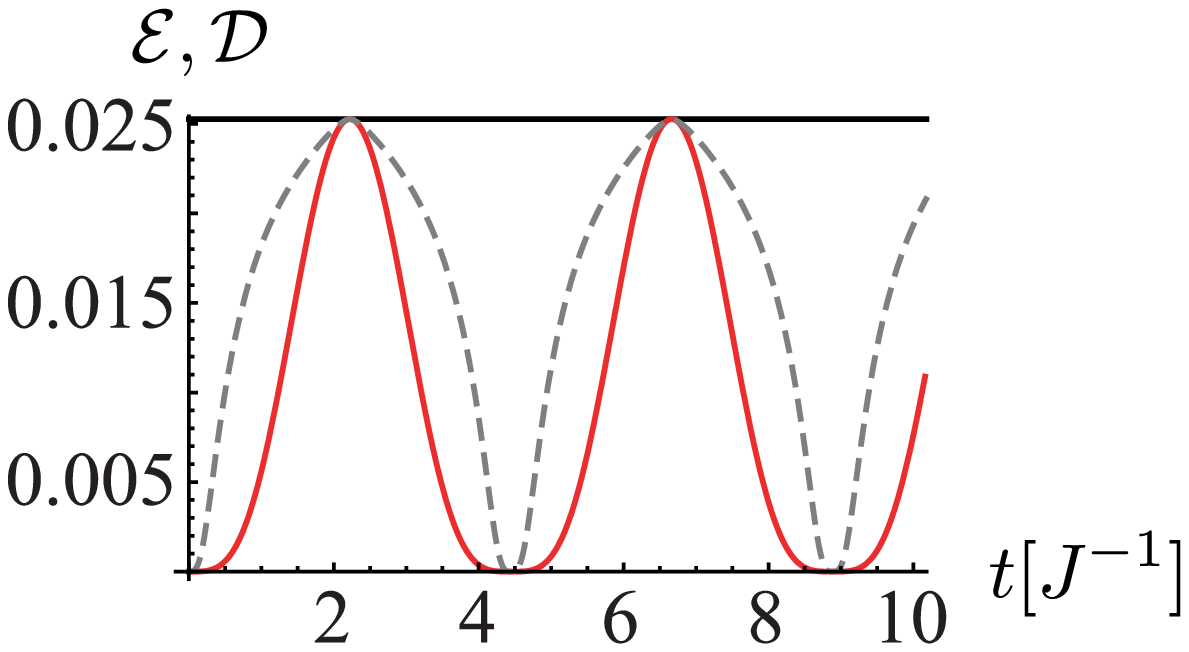}~~\includegraphics[width=0.5\linewidth]{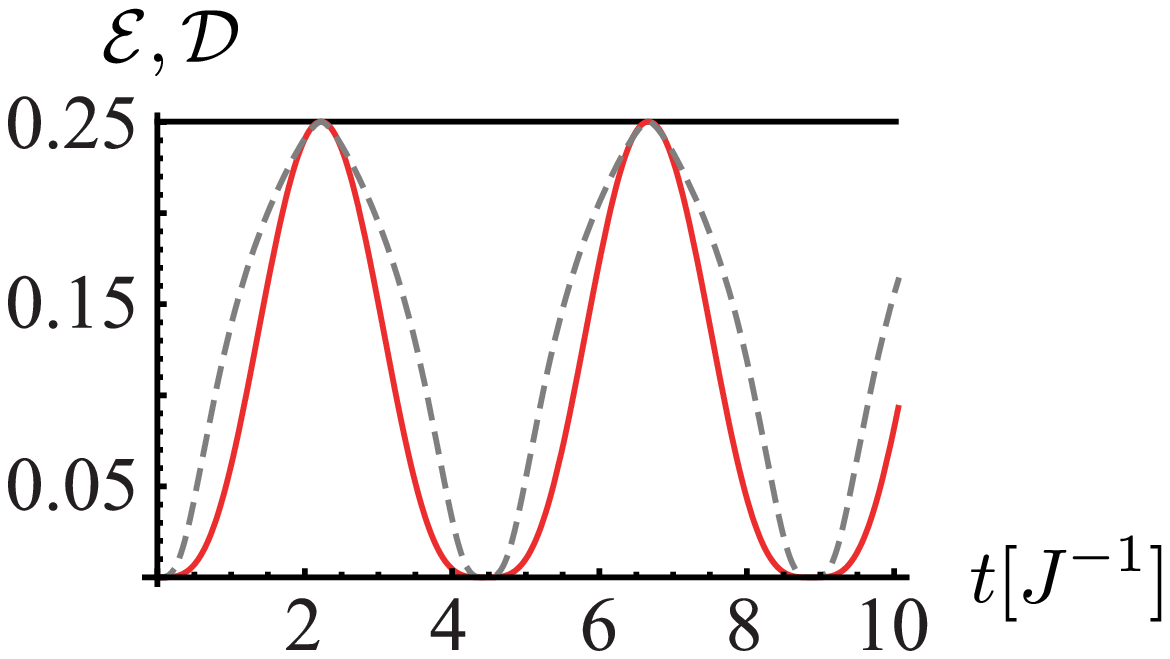}\\
{\bf (c)}\hskip3cm{\bf (d)}
\includegraphics[width=0.5\linewidth]{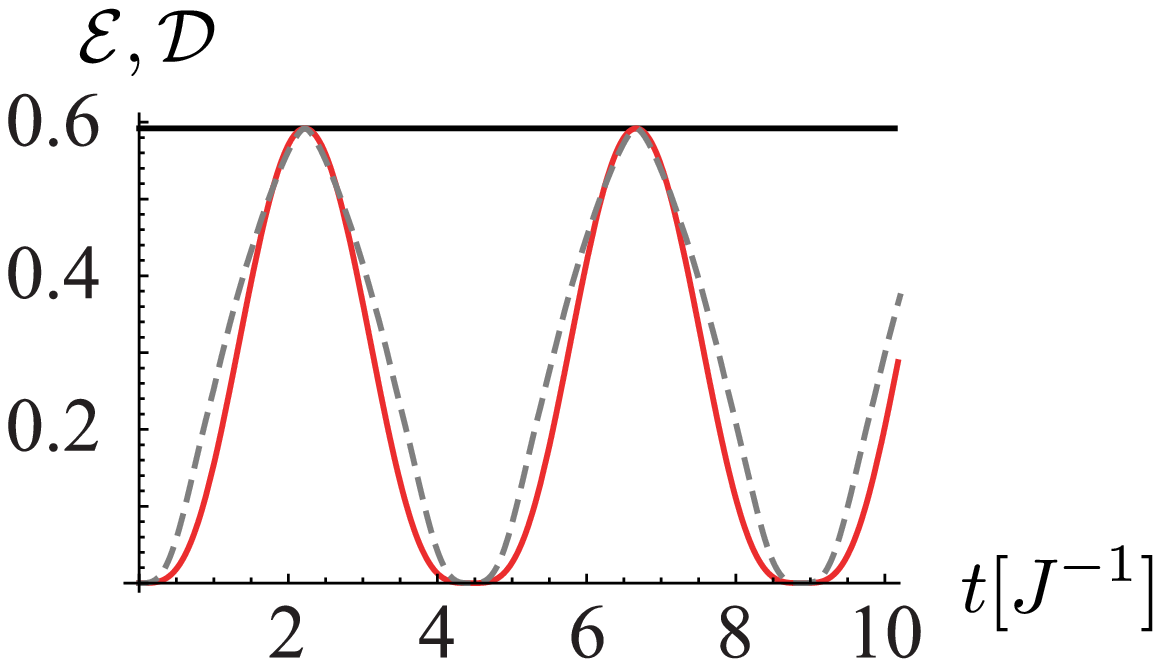}~~\includegraphics[width=0.5\linewidth]{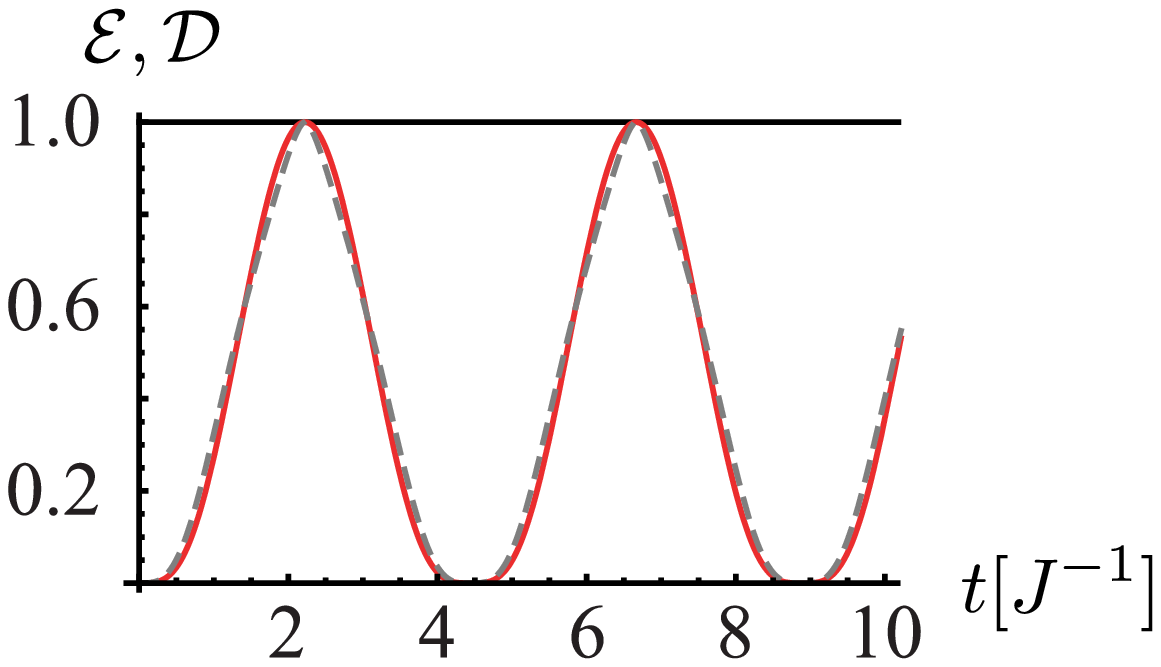}
\caption{(Color online) Time behavior of quantum correlations
between spins $0$ and $3$ in a system ruled by
$\hat{\cal H}$ with uniform interaction strengths and arbitrary $h$. 
The panels (a)-(d) are for $C_{(1,0)}=0.1, 0.4, 0.7,
1$ respectively, and the initial EoF is shown as a straight line.
The solid (red) curve shows the EoF, while the dashed (gray) one
is for the shared two-way QD.} \label{figure1}
\end{figure}

\begin{figure}[b]
{\bf (a)}\hskip3cm{\bf (b)}
\includegraphics[width=0.5\linewidth]{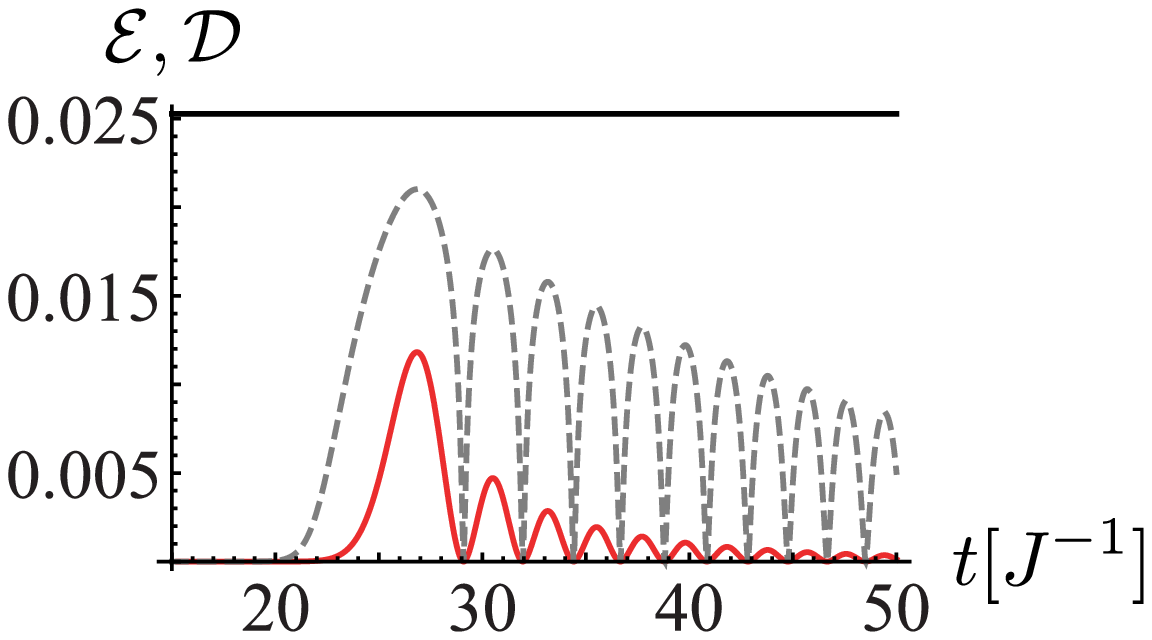}~~\includegraphics[width=0.5\linewidth]{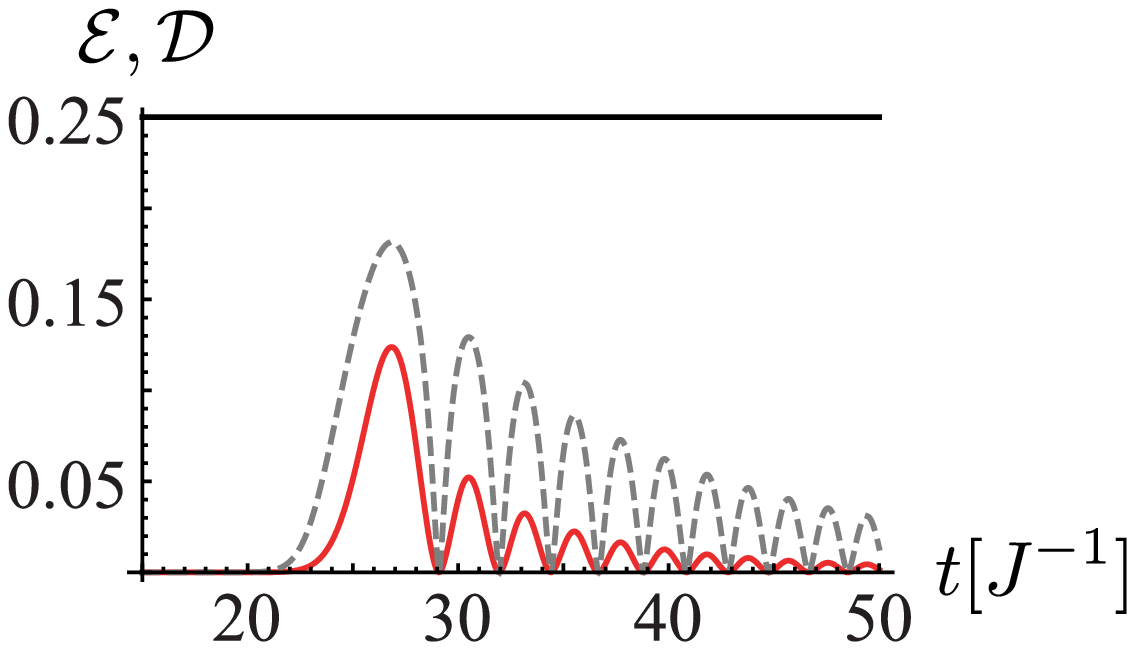}\\
{\bf (c)}\hskip3cm{\bf (d)}
\includegraphics[width=0.5\linewidth]{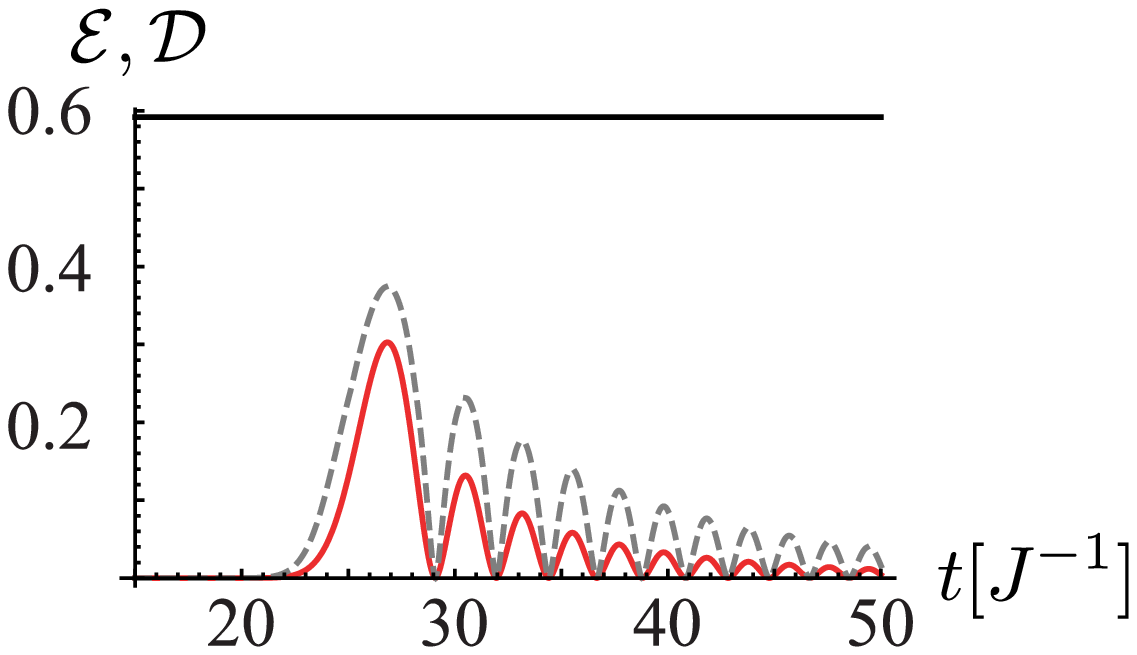}~~\includegraphics[width=0.5\linewidth]{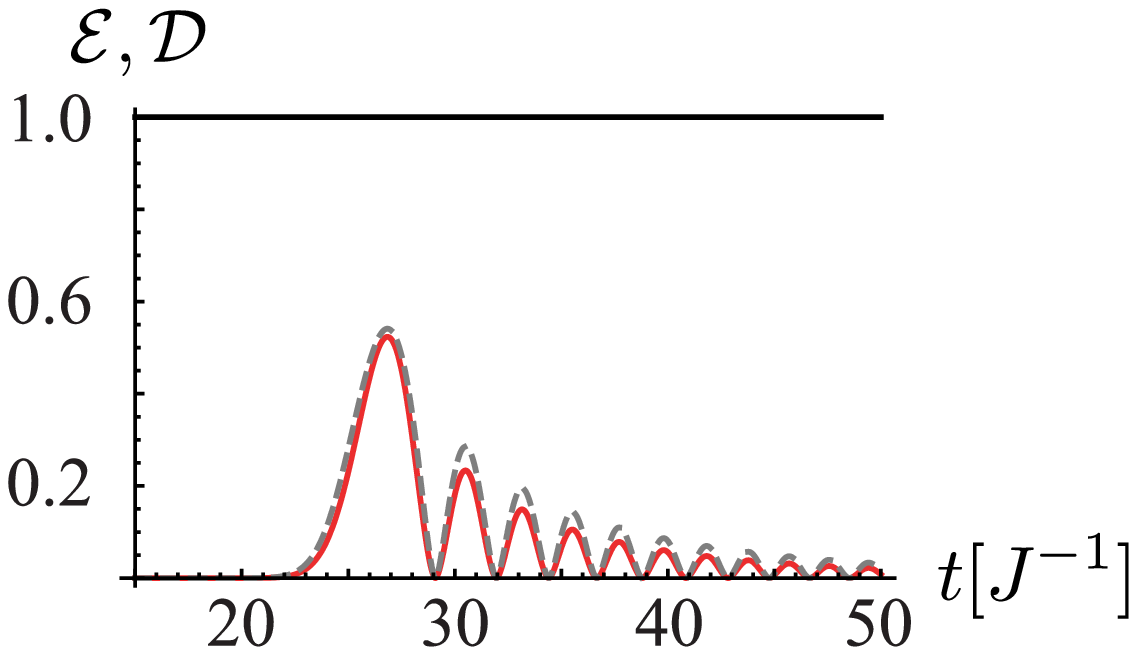}
\caption{(Color online) Time behavior of quantum correlations
between spin $0$ and $50$ for a system ruled by
$\hat{\cal H}_T$. with arbitrary $h$. The panels (a)-(d) are for
$C_{(1,0)}=0.1, 0.4, 0.7, 1$ respectively, and the initial EoF is
shown as a straight line. The solid (red) curve shows the EoF
${\cal E}$, while the dashed (gray) one is for the shared two-way
QD ${\cal D}$.} \label{figure2}
\end{figure}

The behavior disclosed above becomes much more visible when we
consider longer chains. The uniformity of the coupling strengths
across the medium prevents perfect state transfer (and thus
perfect entanglement transfer) for chains of more than three
spins~\cite{kay,cambridge}. Therefore, we find an interesting
feature in the corresponding propagation of ${\cal D}$ [see
Fig.~\ref{figure2}]: the discrepancy between the propagated EoF
and QD becomes quantitatively  much more significant, while the
degree of propagated discord is damped in time much more slowly
than entanglement. However, such effects depend strongly on the
initial value of entanglement in a way that the behavior of ${\cal
E}$ and ${\cal D}$ almost merge as ${\cal E}_{(1,0)}\rightarrow1$.
In Fig.~\ref{figure2} we show an instance of this case by
reporting the propagation of ${\cal E}$ and ${\cal D}$ across a
system of 50 spins interacting according to $\hat{\cal H}$.

Needless to say, as the two figures of merit refer to two
different forms of quantum correlations, some quantitative
differences should be expected. However, here we would like to
stress that, the input state pure, as remarked above, QD and
EoF are {exactly} equivalent. This implies that the differences
$|{\cal D}(t)-{\cal E}(0)|$ and $|{\cal E}(t)-{\cal E}(0)|$
faithfully quantify the performance of each non-classicality
indicator upon propagation. 
%The smallest between the two propagates better.
\begin{figure*}[t]
{\bf (a)}\hskip5cm{\bf (b)}\hskip5cm{\bf (c)}
\includegraphics[width=0.3\linewidth]{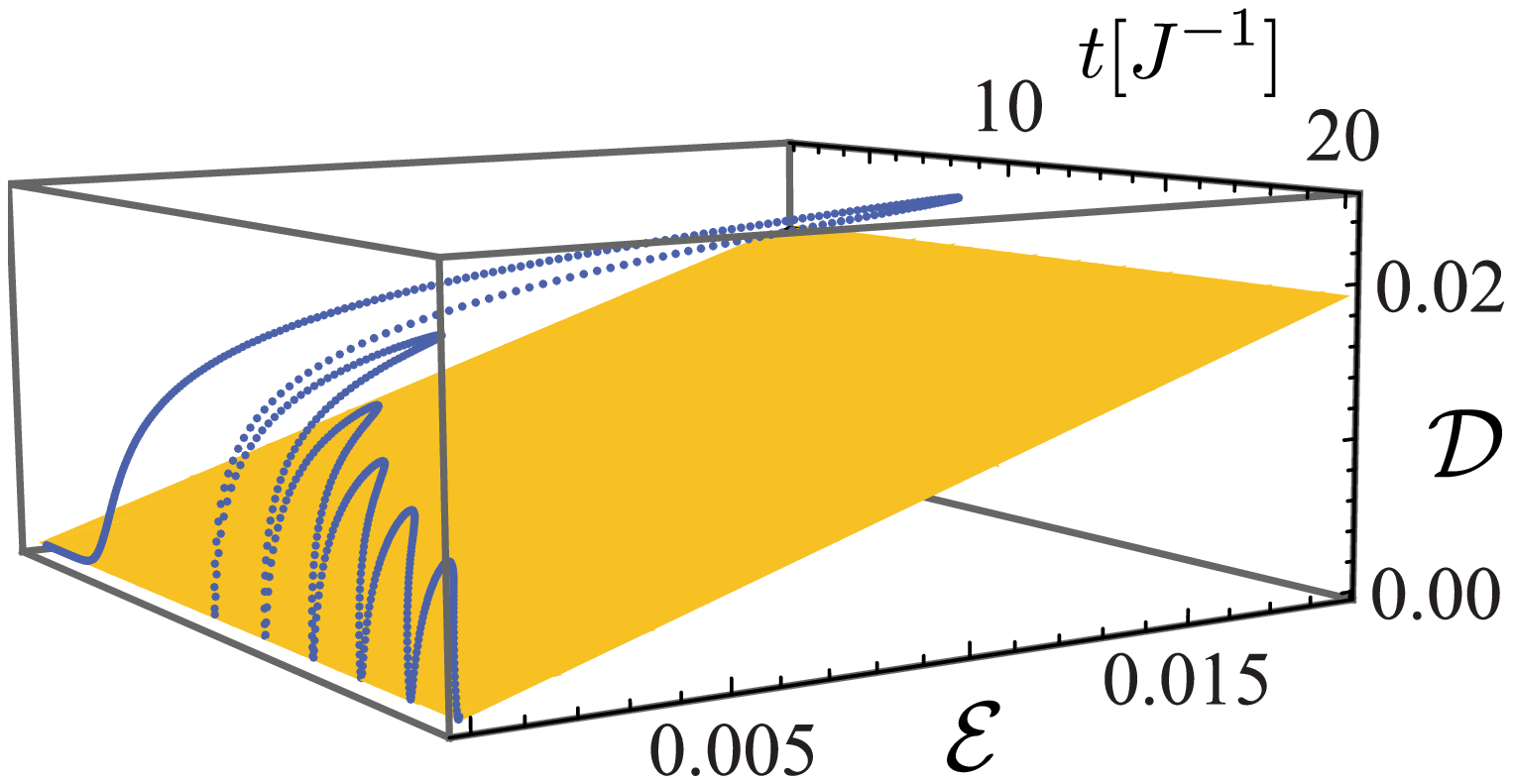}~~\includegraphics[width=0.3\linewidth]{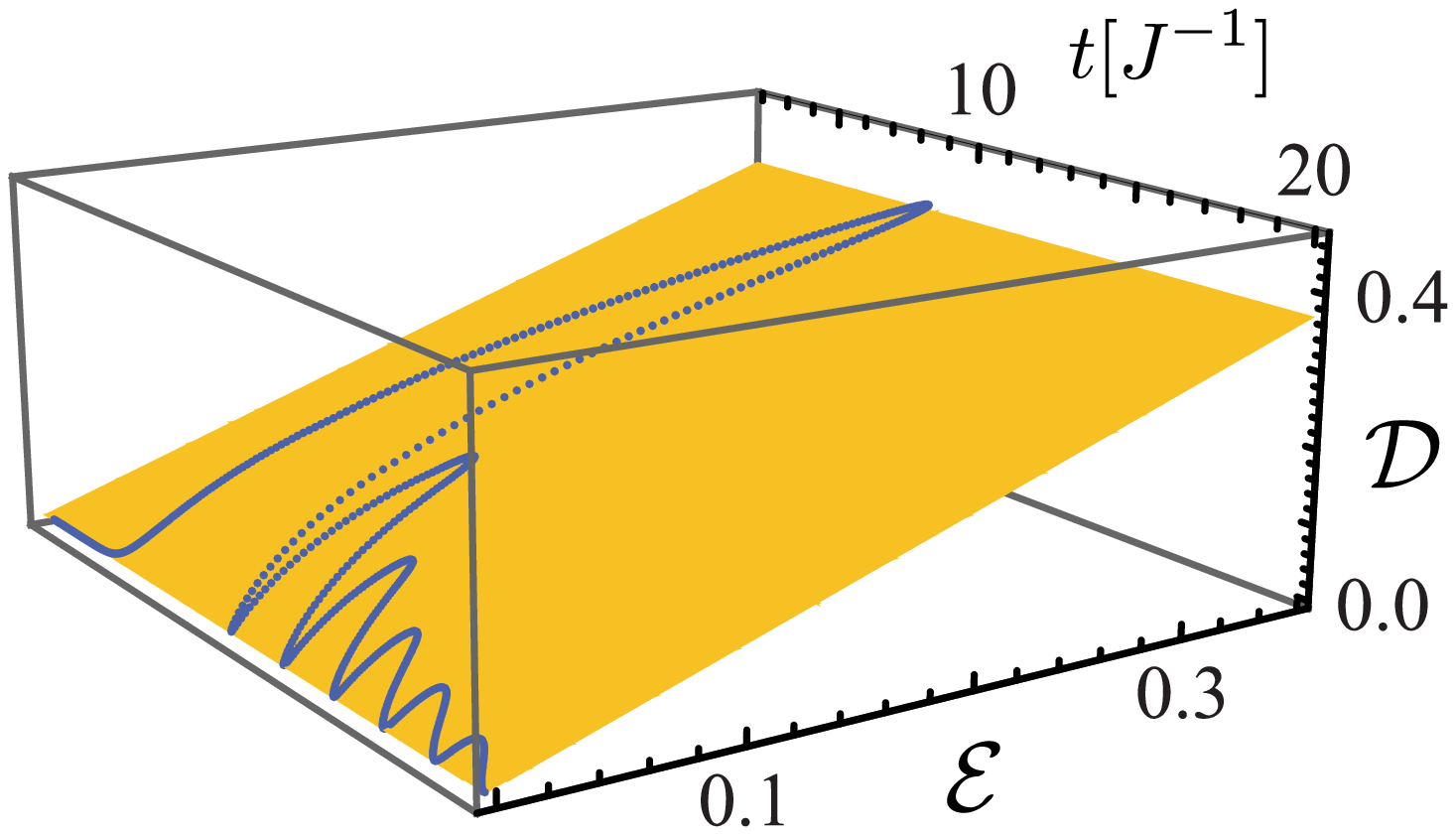}\includegraphics[width=0.3\linewidth]{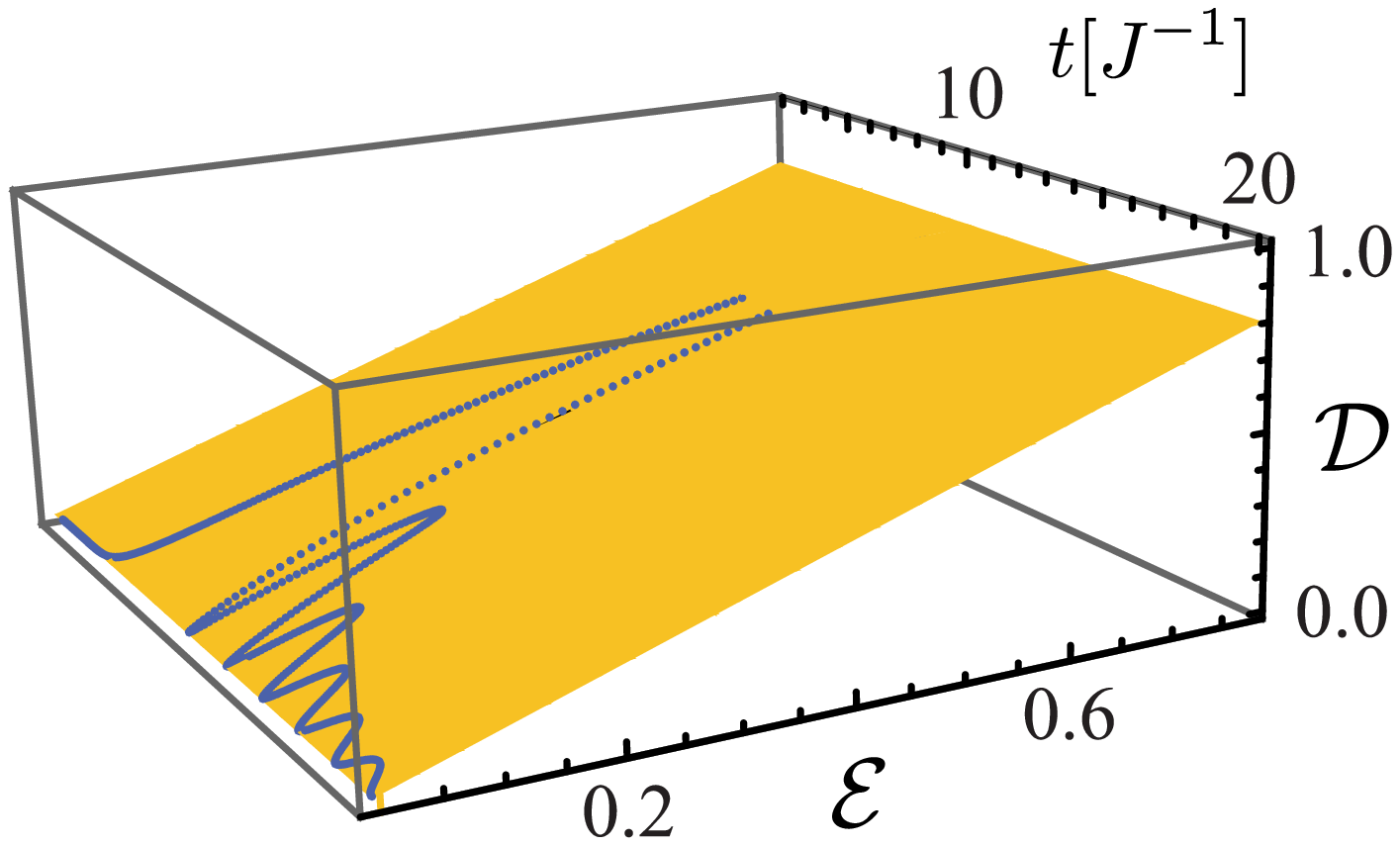}
\caption{(Color online) Behavior of QD against entanglement and
propagation time for a chain of 15 spins, homogeneous intra-chain
couplings and arbitrary $h$. The yellow plane at ${\cal D}{=}{\cal
E}$ is used as a
guide to the eye for discerning whether or not ${\cal D}{\ge}{\cal
E}$. Each panel is for a different initial value of concurrence in
pair $(1,0)$. We have taken $C_{(1,0)}\,{=}\,0.1,0.6$ and 1 in
panel {\bf (a)}, {\bf (b)} and {\bf (c)}, respectively. As the
initial degree of entanglement grows, the efficiency of transport
of ${\cal D}$ gets very close to that for ${\cal E}$.}
\label{figure3}
\end{figure*}

These issues are better discussed by looking at
Fig.~\ref{figure3}, where we plot the propagated QD and EoF
against time for a total of 15 spins. In each panel, the plane for
which ${\cal E}(t)\,{=}\,{\cal D}(t),~\forall{t}$ is displayed as
a reference. Any point lying above the plane corresponds to
discord overcoming EoF. Our choice of $N\,{=}\,15$ is only due to
the clarity of the associated figures, and does not hinder the
validity of our conclusions.

\begin{figure}[b]
\hskip0.2cm{\bf (a)}\hskip2.5cm{\bf (b)}\hskip2.5cm{\bf (c)}
\includegraphics[width=1.01\linewidth]{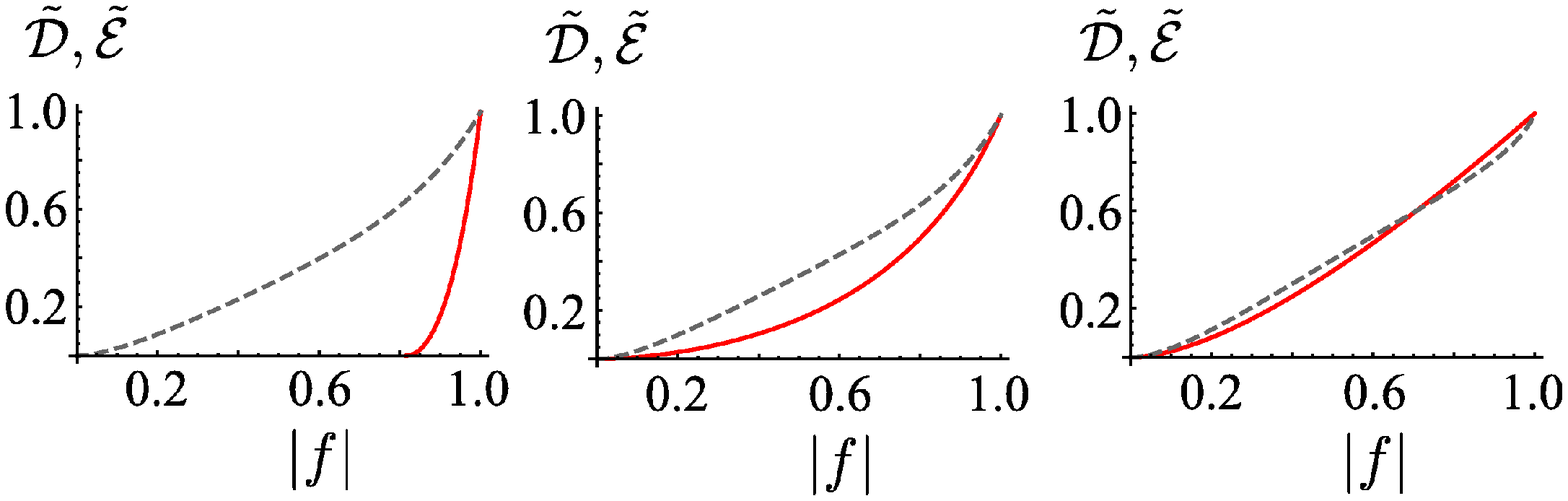}
\caption{(Color online) Comparison between the re-scaled
quantities $\tilde{\cal D}$ (grey dashed line) and $\tilde{\cal E}$
(red solid one) propagated across our chain of uniform couplings and
arbitrary magnetic field. The spin-pair $(1,0)$ is prepared in a Werner
state with $a\,{=}\,0.4,0.7$ and $1$ [panel {\bf (a)}, {\bf (b)}
and {\bf (c)} respectively]. The curves are plotted against the
transition amplitude $|f|$.} \label{figureW}
\end{figure}

We now show that the qualitative features revealed by our study on
pure input states hold also when a mixed state of the spin-pair
$(1,0)$ is prepared. As in general the value of discord and
entanglement associated to such input states will not coincide,
this situation encompasses from the start the profound differences
between QD and EoF. In order to wash out the ambiguities
associated with the possible choices for input mixed states, we
refer to the studies in Refs.~\cite{alqasimi, MNCMS}. There,
families of two-spin mixed states maximizing the degree of the
two-way QD at fixed values of the global entropy have been
identified and fully characterized. Furthermore, for these states
$\mathcal{D}^\leftarrow\,{=}\,\mathcal{D}^\rightarrow$ so it
is immaterial which spin is attached to the chain. We dub such
states as maximally discorded mixed states (MDMS), which represent
the counterpart, as far as discord is concerned, of the well-known
maximally entangled mixed states (MEMS)~\cite{MEMS}. Even more
interesting, under proper choices of entanglement measures, part
of the MEMS frontier is shared with the MDMS one~\cite{MNCMS}. As
such extremal states are clearly dressed with a particular
significance, we now restrict our study to them. The MDMS boundary
is a patch-work of three classes of $X$-type states whose
elements, following the notation used in Eq.~(\ref{XXX}), are
given by
\begin{equation}
\begin{aligned}
\rho^P_{11}&=\rho^P_{14}=\frac{a}{2},~~~~\rho^P_{22}=\frac{1-a-g}{2},~~~~\rho^P_{33}=\frac{1-a+g}{2},\\
\rho^W_{11}&=%\varrho^W_{44}=
\frac{1+a}{4},~~~~\rho^W_{22}=\rho^W_{33}=\frac{1-a}{4},~~~~\rho^W_{14}=\frac{a}{2},\\
\rho^R_{11}&=\frac{1-a}{2},~~~~\rho^R_{22}=a,~~~~\rho^R_{33}=0,~~~~\rho^R_{14}=\frac{g}{2}
\end{aligned}
\end{equation}
and $\rho^{P,W,R}_{23}\,{=}\,0$. While $\rho^P$ is a general
two-parameter family ($a+g\,{\le}\,{1}$), $\rho^W$
($-1/3\,{\le}\,{a}\,{\le}\,{1}$) are Werner states~\cite{werner}
and $\rho^R$ (${0\,{\le}\,{a}\,{\le}\,{1/3}}$ with
$a+g\,{\le}\,1$) are MEMS when the relative entropy is chosen as a
measure of entanglement~\cite{MEMS}. Such states belong to the
MDMS frontier only under properly chosen values of $a$ and $g$.
Such conditions are in general highly non-trivial and we refer to
Ref.~\cite{MNCMS} for full details. Here it is enough to state
that we will consider values of such
parameters that guarantee the MDMS nature of the corresponding
states.

In order to provide a faithful evaluation of their performance
upon propagation, we will compare the re-scaled quantities
$\tilde{\cal R}\,{=}\,{\cal R}/{\cal R}_{(1,0)}$ with ${\cal
R}\,{=}\,{\cal E}, {\cal D}$ [${\cal R}_{(1,0)}\,{=}\,{\cal
E}_{(1,0)}, {\cal D}_{(1,0)}$] being the value of one of our
figures of merit after propagation [for the initial spin pair].
Moreover, rather than replicating the time-dependent study
performed so far and in order to provide a {\it universal}
analysis freed from the choice of $N$, we will consider the
propagated QD and EoF as general functions of the
single-excitation transition amplitude $|f|\in[0,1]$ (from now on
we drop the label stating its dependence on time).

We start by studying Werner states, which are entangled only for
$a\,{\ge}\,1/3$. For values mildly larger than this threshold,
where the purity of the state is small and also its entanglement,
very large values of $|f|$ are required in order to actually
transport ${\cal E}$. 
% Under dynamical conditions that guarantee
% $|f|\,{=}\,1$, the whole initial EoF is transported, regardless of
% ${\cal E}_{(1,0)}$. 
Differently, ${\cal D}$ is non-null for any
$|f|$ and irrespectively of the initial QD properties of pair
$(1,0)$. The relative discrepancy between the two figures of merit
is in general very large and decreases only for almost ideal transport of
excitations across the chain. As $a\rightarrow1$, i.e., by
increasing the purity of the state, more EoF is transported, even
in the low-$f$ region, thus reducing the differences between the
two non-classicality indicators. In the limit of $a\,{=}\,1$,
which makes $\rho^W$%\,{=}\,\ket{\psi_P(0)}\!\bra{\psi_P(0)}$,
a maximally entangled pure state, discord is overtaken by the EoF at $|f|\,{\ge}\,1/\sqrt 2$. In
fact, the state of pair $(N,0)$ corresponding to such a value of the
transition amplitude reads
\begin{equation}
\tilde\rho^W=\left[
\begin{matrix}
\frac{1}{2}&0&0&\frac{1}{2\sqrt 2}\\
0&\frac{1}{4}&0&0\\
0&0&0&0\\
\frac{1}{2\sqrt 2}&0&0&\frac{1}{4}
\end{matrix}
\right],
\end{equation}
which is an instance of a mixed state (its von Neumann entropy is
$0.811278$) with ${\cal D}\,{=}\,{\cal E}$. Fig.~\ref{figureW}
exemplifies the cases discussed above for three different values
of parameter $a$.

\begin{figure}[b]
\includegraphics[width=0.65\linewidth]{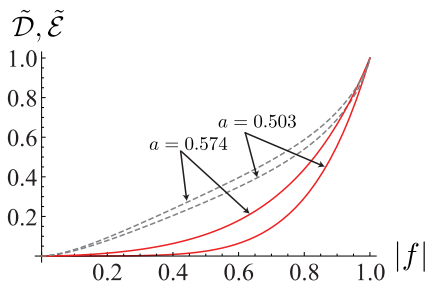}
\caption{(Color online) Comparison between the re-scaled
quantities $\tilde{\cal D}$ (grey dashed line) and $\tilde{\cal E}$
(red solid one) propagated across our chain of uniform couplings and
arbitrary magnetic field. The spin-pair $(1,0)$ is prepared in state
$\rho^P_{(1,0)}$ with $b\,{=}\,0$ and two values of $a$. The
curves are plotted against the transition amplitude $|f|$.}
\label{figure2P}
\end{figure}

\begin{figure}[t]
\hskip0.2cm{\bf (a)}\hskip3.5cm{\bf (b)}
\includegraphics[width=1.\linewidth]{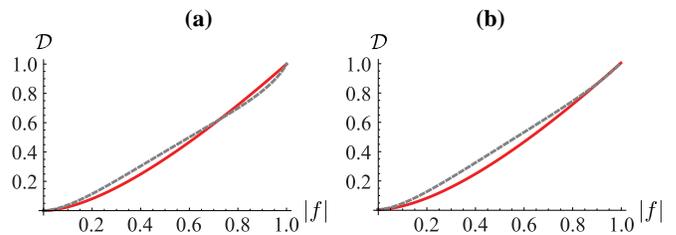}
\caption{(Color online) Comparison between the rescaled quantities
$\tilde{\cal D}$ (grey dashed line) and $\tilde{\cal E}$ (red solid one)
propagated across our chain of uniform couplings and no magnetic
field. The spin-pair $(1,0)$ is prepared in state $\rho^R_{(1,0)}$
with $(a,g)\,{=}\,(0.0150,0.9747)$ [panel {\bf (a)}] and
$(a,g)\,{=}\,(0.1625,0.7649)$ [panel {\bf (b)}]. The curves are
plotted against the transition amplitude $|f|$.} \label{figureR}
\end{figure}

When addressing the case of the two-parameter family $\rho^P$, the
situation is even more striking. This class of states is MDMS for
$g\,{=}\,0$ and $a\,{\in}\,[0.503,0.574]$, for instance, which
correspond to a region of large mixedness such that the re-scaled
QD is always larger than the corresponding re-scaled EoF, as seen
in Fig.~\ref{figure2P}. In this case, $\tilde{\cal E}$ never
overcomes $\tilde{\cal D}$ and can only equal it at $|f|\,{=}\,1$.

Finally, the general picture is confirmed by the investigation on
the third family of boundary states, $\rho^R$, which  are also the
states maximizing the relative entropy of entanglement at fixed
global mixedness~\cite{MEMS}. The conditions that $a$ and $g$
should satisfy in order for $\rho^R$ to be a MDMS are rather
non-trivial, passing through the solution of a transcendental
equation~\cite{MNCMS}. For our purposes, it is enough to state
that $\rho^R$ spans the large-purity region of the boundary and
thus the crossing of QD and EoF at some value of $|f|$ should be
expected. This is indeed the case, as highlighted in
Fig.~\ref{figureR} for two values of parameters $a$ and $g$ that
guarantee the MDMS nature of such class of states. While
Fig.~\ref{figureR} {\bf (a)} refers to a large-purity case (the
von Neumann entropy of the corresponding state is $0.159$, implying a small degree of mixedness) where EoF wins over
QD at large enough $|f|$, panel {\bf (b)} is for a much more mixed
state. In this case, as expected from the analysis above, ${\cal
D}$ is better transported at any dynamical condition.

\subsection{Effects of a magnetic field}
\label{Ss.magnetic} Here we study the effects of a uniform
magnetic  field $h$ on the transport of quantum correlations. From
Eq.~(\ref{E.amplitude}) we see that the single-excitation
transition amplitude at $h\,{\neq}\, 0$ differs from that at
$h\,{=}\,0$ only by an overall oscillating phase factor,
 $f^{(h)}(t)\,{=}\, e^{- i 2 h t} f^{(0)}(t)$.
Consequently, by using the general form of an evolved state in Eq.~(\ref{e.rhooutrr'})
and the expression for the concurrence of $X$-type states given in Eq.~(\ref{e.conc2}),
it is straightforward to see that ${\cal E}$ would depend on just $|f(t)|$
so that the introduction of a magnetic field does not affect the propagation of EoF.

The situation is radically different when considering the
transport of QD for a certain class of input states of pair
$(1,0)$. In fact, by considering states with maximally mixed
marginals~\cite{Luo2008}
$\rho^{(1,0)}(0)\,{=}\,\frac{1}{4}\left(\hat\openone+\sum_{i=x,y,z}
c_i \hat\sigma_i\otimes\hat\sigma_i \right)$ ($c_i\in\mathbb{R}$), 
for which analytic expressions for QD hold, we find that the
magnetic field {increases} the amount of discord that can be
obtained between $0$ and $N$ as compared to the case with
$h\,{=}\,0$. The reason for this enhancement can be found in
the fact that, when $|c_{x}|\,{\neq}\,|c_{y}|$, ${\cal D}$ depends
on both the real and imaginary part of $f$. Moreover, the phase
factor $e^{- i 2 h t}$ in  $f^{(h)}(t)$ yields an oscillating behavior of ${\cal D}$
that is lower-bounded by the value achieved at $h\,{=}\,0$. In
Fig.~\ref{mag} we show the typical behavior described above.

$X$-type states with maximally mixed marginals allow for the
identification of cases where the conditions for a truthful
transport are breached. In all the cases studied so far, QD is a
monotonically increasing function of $|f|$ with a maximum
occurring at $t^* \sim N$, which are all features consistent with
the picture of a transmission mechanism. For a
maximally-mixed-marginal state having $\left|c_x\right|\,{\neq}\,
\left|c_y\right|$, on the other hand, the discord between spin $0$
and $N$ can be larger than the initial QD content of $\rho^{(1,0)}$.
This point is best illustrated with the aid of an example. The
density matrix elements of a two-spin X-type state with zero
discord should satisfy one of the following
conditions~\cite{Li2011,MNCMS}
\begin{enumerate}[{\bf 1.}]
\item $\rho_{14}\,{=}\,\rho_{23}\,{=}\,0$. In this case, all the
coherences are identically null and $\rho$ is purely diagonal. The
corresponding state is thus a classical-classical one as
$\mathcal{D}^{\leftarrow,\rightarrow}=0$. \item
$\rho_{11}=\rho_{22}$, $\rho_{33}=\rho_{44}$ and
$\left|\rho_{14}\right|=\left|\rho_{23}\right|$. This case
corresponds to a quantum-classical state with
$\mathcal{D}^\rightarrow\,\neq\, 0$ and $\mathcal{D}
^\leftarrow\,{=}\,0$. \item $\rho_{11}=\rho_{33}$,
$\rho_{22}=\rho_{44}$ and
$\left|\rho_{14}\right|=\left|\rho_{23}\right|$, which give rise
to a classical-quantum state with $\mathcal{D}^\rightarrow\,{=}\,
0$ and $\mathcal{D} ^\leftarrow \,\neq\, 0$.
\end{enumerate}
When the symmetrized discord in Eq.~(\ref{e.Discord}) is used as a
figure of merit, conditions {\bf 2} and {\bf 3} collapse into
\begin{enumerate}[{\bf 4.}]
\item $\rho_{ii}{=}{1}/{4}~(i{=}1,...,4)$ with
$\left|\rho_{14}\right|\,{=}\,\left|\rho_{23}\right|$.
\end{enumerate}
We now consider a $\rho^{(1,0)}(0)$ having $c_y\,{=}\,c_z\,{=}\,0$
and $c_x\,{=}\,1$, which corresponds to a state endowed with only
classical correlations (embodied by
$\langle\hat{\sigma}^{(1)}_x\otimes\hat{\sigma}^{(0)}_x\rangle\neq0$). The evolution
yields a quantum-classical state
\begin{equation}
\rho^{(N,0)}=
\begin{pmatrix}
\frac{2-\left|f\right|^2}{4}&0&0&\frac{f}{4}\\
0&\frac{2-\left|f\right|^2}{4}&\frac{f}{4}&0\\
0&\frac{f}{4}&\frac{\left|f\right|^2}{4}&0\\
\frac{f}{4}&0&0&\frac{\left|f\right|^2}{4}
\end{pmatrix}.
\end{equation}
If $f$ is a real (imaginary) function, such state develops a non-zero correlation function  
$\langle\hat{\sigma}^{(N)}_x\,{\otimes}\,\hat{\sigma}^{(0)}_x\rangle$ 
($\langle\hat{\sigma}^{(N)}_y\,{\otimes}\,\hat{\sigma}^{(0)}_x\rangle$) that is quantitatively equal to $f$ itself. However, 
 spin $N$ also develops a non-zero magnetization in the $z$-direction 
given by $\langle\hat{\sigma}^{(N)}_z\rangle\,{=}\,1-\left|f\right|^2$.
According to the results in
Refs.~\cite{Ferraro2010,Dakic2010}, due to the non-commutativity
between $ \hat{\sigma}^{(N)}_z\,{\otimes}\,\hat\openone^{(0)}$ and
$\hat{\sigma}^{(N)}_x\,{\otimes}\,\hat{\sigma}_x^{(0)}$, the initially
classical-type correlations acquire a quantum nature responsible
for non-zero discord, as is shown in Fig.~\ref{F.magnetic}.
Reasoning along the same lines, we can see that an initially
zero-QD state fulfilling condition {\bf 1} remains such because
all one- and two-spin correlators involve only $\hat\sigma_z$.

\begin{figure}[ht]
\includegraphics[width=0.5\textwidth]{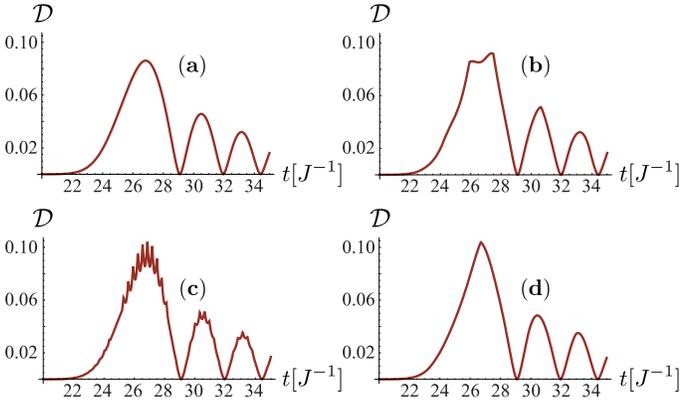}
\caption{(Color online) QD transported across a chain of $50$
spins whose $(1,0)$ pair is initialized in an $X$-type state with
maximally mixed marginals having $c_x{\simeq}0.53$,
$c_y{\simeq}0.340$, $c_z{\simeq}0.035$. The associated value of QD
is ${\cal D}_{(01)}(0)\,{=}\,0.210$. We have taken $h/J\,{=}\,0$
[panel ${\bf (a)}$], $h/J\,{=}\,0.5$  [panel ${\bf (b)}$],
$h/J\,{=}\,1$ [panel ${\bf (c)}$] and $h/J\,{=}\,5$ [panel ${\bf
(d)}$].} \label{mag}
\end{figure}

\begin{figure}
\includegraphics[width=0.75\linewidth]{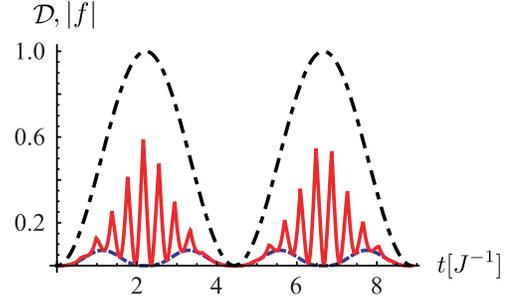}
\caption{(Color online) QD between spin $0$ and $3$, for a chain
of N=$3$ spins whose pair $(1,0)$ is initialized in an $X$-type
state with maximally mixed marginals having $c_x\,{=}\,1$,
$c_y\,{=}\,c_z\,{=}\,0$ and thus ${\cal D}(0)=0$. We have taken
$h/J\,{=}\,0$ (blue dashed line) and $h/J\,{=}\,2$ (full red
line). The dot-dashed line is the transition amplitude
$\left|f_{3}(t)\right|$. } \label{F.magnetic}
\end{figure}

\subsection{Looking at the spin-chain as a channel}
\label{Ss.environment} We now slightly change perspective and
consider the spins occupying sites $j\,{=}\,2,...,N$ as the
elements of an {\it environment} for spin $1$. This allows us to
investigate the robustness of the quantum correlations shared with
spin $0$ under the influence of an amplitude damping channel. A
similar analysis can be found in Ref.~\cite{Apollaro2010} where
the entanglement dynamics in the presence of the same model has
been studied. Here we extend this study to the analysis of QD so as to show that, at variance with entanglement, 
a unilateral non-unitary channel can induce quantum correlations in an initial state that is fully classical.

By using Eqs.~(\ref{e.rhooutrr'}) and (\ref{e.conc2}), we obtain
that entanglement sudden death (ESD) occurs when
$\left|f_{1}(t)\right|^2\,{\leq}\,
1\,{-}\,\frac{\left|\rho_{14}\right|^2-\rho_{22}\rho_{33}}{\rho_{33}\rho_{44}}$
and $\left|f_{1}(t)\right|^2\,{\leq}\,
1\,{-}\,\frac{\left|\rho_{23}\right|^2-\rho_{11}\rho_{44}}{\rho_{33}\rho_{44}}$,
for $\rho_{33}\rho_{44}\,{\neq}\, 0$. For states with initial
finite entanglement, the quantities
$\left|\rho_{14}\right|^2-\rho_{22}\rho_{33}$ and
$\left|\rho_{23}\right|^2-\rho_{11}\rho_{44}$ are always comprised
between $0$ and $1$. For $\rho_{33} \left(\rho_{44}\right)\,{=}\,
0$, the concurrence evolves as
$C_{(1,0)}(t)\,{=}\,\left|f_{1}(t)\right| C_{(1,0)}(0)$. Besides this
latter case, where ESD occurs only at
$\left|f_{1}(t)\right|\,{\equiv}\, 0$, mixed
non-separable quantum states experience ESD due to the fact that
appropriate choices of the interaction parameters make
$\left|f_{1}(t)\right|$ range from 0 to 1: The amplitude damping
channel, acting only on one spin, is able to erase completely the
quantum correlations which give rise to entanglement.

Now let us turn our attention to the conditions for the vanishing
of QD discussed previously. As far as condition {\bf 1} is
concerned, we observe that a zero-QD state will remain such at all
times because an environment addressing only one spin of a
bipartite state cannot build up quantum coherences. If the latter
are initially present in such a way to start from a non-zero QD
state, condition {\bf 1} can be possibly fulfilled only
asymptotically. On the other hand, starting from a zero-QD state
according to condition {\bf 4}, we note that the constraint
$\rho_{ii}=1/4$ breaks down because of the dynamics embodied by
Eqs.~(\ref{e.rhooutrr'}) and the classical-classical state evolves
into a quantum-classical one, so that $\mathcal{D}
^\rightarrow\,{\neq}\, 0$. Conversely, starting from a state with
non-zero QD, it is necessary, in order to satisfy condition {\bf 4},
that the initial state consists of a quantum-classical state (condition {\bf 2}) and
that the environment acts on the spin that, when subjected to a
complete projective measurement, gives rise to non-zero one-way
QD. Furthermore, as the amplitude damping channel implies
$\rho_{11}(t)\,{\geq}\, \rho_{11}(0)$, only quantum-classical
states with $\rho_{11}(0)\,{<}\,\frac{1}{4}$ can evolve to a
classical-classical one. This will occur at times $t^*$ such that
$f_{1}(t^*)={1}/{2}+\rho_{11}/(1-2 \rho_{11})$.

\section{Conclusions}
\label{conc} We have studied the propagation of quantum
correlations across a chain of interacting spins by looking at the
performance of two significant figures of merit: quantum discord
and entanglement of formation. The amount of transported quantum
correlations has been quantified when the chain is seeded with
various instances of two-spin mixed states. We have explicitly
considered the case of pure entangled states, as well as the
members of the boundary family of MDMSs, which maximize QD at set
values of global entropy. Discord appears to be consistently
better transported than entanglement, except for highly pure input
states whose EOF is transported across a chain guaranteeing a
large single-excitation transition amplitude. Moreover, we have
performed a brief case-study on the conditions under which QD is
actually created upon propagation across the chain, pointing
towards the class of states giving rise to this effect, and
analyzing, in particular, how the discord increases under the
effect of a magnetic field. Finally we have revealed that, whereas
the EoF of all mixed states undergoes ESD by an appropriate choice
of the interaction parameters, QD vanishes only under very
specific initial state conditions and interaction settings. In
particular quantum-quantum states allow
for vanishing discord under the influence of the {\it spin
environment} considered here only asymptotically in time.

Our work extends the investigations performed so far on the propagation of entanglement 
to the broader realm of more general quantum correlations proving that,
 in the dispersive medium consisting of interacting quantum particles, correlations beyond entanglement are favored. 
Such results motivate the study on a few aspects of this general problem that still remain to be addressed, 
such as the extension to other quantifier of non-classicality, 
their behavior under more general dynamical conditions of the spin media 
and the quantification of the corresponding non-Markovian effects brought about by the spin chain on the dynamics of general indicators of quantumness.

\acknowledgments

%The authors would like to thank J. Alba's generosity.
We thank Marco Piani and  Paola Verrucchi for discussions. TJGA is grateful to the Centre for Theoretical Atomic, Molecular and Optical Physics at the School of Mathematics and Physics, Queen's University Belfast for hospitality during completion of this work. We acknowledge financial support from the Department of Employment and Learning, the Italian Ministry of Education, University, and Research through the 2008 PRIN program (Contract No.2008PARRTS003), the Irish Research Council for Science, Engineering and Technology and the UK EPSRC (EP/G004579/1). MP and FP acknowledge support by the British Council/MIUR British-Italian Partnership Programme 2009-2010.

%{\it Note added.-} During completion of this work we became 
%aware of the related manuscript by F. Ciccarello and V. Giovannetti 
%{\it Creating quantum correlations through local non-unitary memoryless channels} 
%that appeared simultaneously to our work.

% \begin{figure*}[h!]
% \includegraphics[width=0.4\linewidth]{jessica1.jpg}~~\includegraphics[width=0.4\linewidth]{jessica2.jpg}
% ~~\includegraphics[width=0.4\linewidth]{jessica3.jpg}~~\includegraphics[width=0.4\linewidth]{jessica4.jpg}
% \end{figure*}

\end{document}